\documentclass{ws-ijmpd}

\begin{document}

\markboth{Ghosh, Garain, Chakrabarti and Laurent}
{Monte-Carlo Simulations in a Two component Flow in presence of Outflow.}

%
\catchline{}{}{}{}{}
%

\title{Monte-Carlo Simulations of Thermal Comptonization Process in a Two Component Accretion Flow
Around a Black Hole in presence of an Outflow}

\author{Himadri Ghosh}

\address{S.N. Bose National Centre for Basic Sciences,\\
JD-Block, Sector III, Salt Lake, Kolkata 700098, India.\\
himadri@bose.res.in}

\author{Sudip K. Garain}

\address{S.N. Bose National Centre for Basic Sciences,\\
JD-Block, Sector III, Salt Lake, Kolkata 700098, India.\\
sudip@bose.res.in}

\author{Sandip K. Chakrabarti\footnote{Also at
Indian Centre for Space Physics, Chalantika 43, Garia Station Rd., Kolkata 700084}}

\address{S.N. Bose National Centre for Basic Sciences,\\
JD-Block, Sector III, Salt Lake, Kolkata 700098, India.\\
chakraba@bose.res.in}

\author{Philippe Laurent}

\address{IRFU, Service d'Astrophysique, Bat. 709 Orme des Merisiers, CEA Saclay, 91191\\
Gif-sur-Yvette Cedex, France, philippe.laurent@cea.fr\\  }

\maketitle

\begin{history}
\received{Day Month Year}
\revised{Day Month Year}
\comby{Managing Editor}
\end{history}

\begin{abstract}

A black hole accretion may have both the Keplerian and the sub-Keplerian components. The 
Keplerian component supplies low-energy (soft) photons while the sub-Keplerian 
component supplies hot electrons which exchange their energy with the soft photons
through Comptonization or inverse Comptonization processes. In the sub-Keplerian flow,
a shock is generally produced due to the centrifugal force. The post-shock region is known as the 
CENtrifugal pressure supported BOundary Layer or CENBOL. 
We compute the effects of the thermal and the bulk motion Comptonization on the soft photons emitted 
from a Keplerian disk by the CENBOL, the pre-shock sub-Keplerian disk and the outflowing jet.
We study the emerging spectrum when both the converging inflow 
and the diverging outflow (generated from the CENBOL) 
are simultaneously present. From the strength of the shock, we calculate the 
percentage of matter being carried away by the outflow and determined
how the emerging spectrum depends on the the outflow rate. 
The pre-shock sub-Keplerian flow was also found to Comptonize the soft photons significantly.
The interplay among the up-scattering and down-scattering effects
determines the effective shape of the emerging spectrum. By simulating several cases with
various inflow parameters, we conclude that 
whether the pre-shock flow, or the post-shock CENBOL or the emerging jet 
is dominant in shaping the emerging spectrum, strongly depends on the geometry of the 
flow and the strength of the shock in the sub-Keplerian flow.
\end{abstract}


\section{Introduction}	
It is well known (Chakrabarti 1990, hereafter C90) that the flow velocity is the same as the velocity of light $c$
as the matter enters through the event horizon. However, the sound speed is never so high.
Thus the incoming flow on a black hole is always supersonic and thus these solutions are likely 
to be most relevant in the study of the physical processes around black holes. As the flow
begins its journey sub-sonically very far away, and becomes supersonic on the horizon, 
the flow is also known as a transonic flow. In the context of the spherical flows, 
Bondi (1952) solution of accretion and Parker (1959) solution of 
winds are clear examples of transonic flows. But they have only one sonic points.
In presence of angular momenta, the flow may have two saddle type sonic points
with a shock in between (C90, Chakrabarti, 1996). The solutions with shocks
have been extensively studied in both the accretion and the winds
even when rotation, heating, cooling etc. are included
(Chakrabarti, 1990, 1996). The study demonstrates that
the accretion and the winds are inter-related -- the outflows are generated from the 
post-shock region. Subsequently, in Chakrabarti
(1999, hereafter C99), Das \& Chakrabarti (1999) and Das et al. (2001), the mass outflow rate was computed 
as a function of the shock strength and other flow parameters.
Meanwhile, in the so-called two component advective flow (TCAF) model of 
Chakrabarti \& Titarchuk (1995) and Chakrabarti (1997), the spectral states were 
shown to depend on the location and strength of the shock. Thus, C99
for the first time, brought out the relationship
between the jets and outflows with the presence or absence of 
shocks, and therefore with the spectral states of a black hole candidates. This
paves the way to study the relative importance between the Compton cloud and the 
outflow as far as emerging spectrum is concerned.

Computation of the spectral characteristics have so far concentrated 
only on the advective accretion flows (Chakrabarti \& Titarchuk, 1995;
Chakrabarti \& Mandal, 2006) and the outflow or the base of the jet 
was not included. In the Monte-Carlo simulations of Laurent \& Titarchuk (2007)
outflows in isolation were used, but not in conjunction with inflows.
In Ghosh, Chakrabarti \& Laurent (2009, hereafter Paper I), 
the results of Monte-Carlo simulations in a setup similar to 
that of Chakrabarti \& Titarchuk (1995) was presented. 
In the present paper, we improve this and obtain the outgoing spectrum in presence of 
both inflows and outflows. We also include a Keplerian disk inside an
advective flow which is the source of soft photons. We show how the spectrum 
depends on the flow parameters of the inflow, such as the accretion rates of the 
two components and the shock strength. 
The post-shock region being denser and hotter, it behaves like the so-called 'Compton cloud'
in the classical model of Sunyaev and Titarchuk (1980). This region is known as the CENtrifugal
pressure supported BOundary Layer or CENBOL. Since the shock location and its strength
depends on the inflow parameters, the variation of the size of the Compton
cloud, and then the basic Comptonized component of the spectrum is thus a function 
of the basic parameters of the flow, such as the specific energy, the accretion rate and the 
specific angular momentum. Since the intensity of soft photons determines the Compton cloud 
temperature, the result depends on the accretion rate of the Keplerian component also. 
In our result, we see the effects of the bulk motion Comptonization (Chakrabarti \& Titarchuk,
1995) because of which even a cooler CENBOL produces a harder spectrum. At the same time, 
the effect of down-scattering due the outflowing electrons is also seen, because of which even 
a hotter CENBOL causes the disk-jet system to emit lesser energetic photons. Thus, the net spectrum
is a combination of all these effects.

In the next section, we discuss the geometry of the soft photon source and the Compton cloud in our
Monte-Carlo simulations. In \S 3, we present the variation of the thermodynamic quantities and other
vital parameters inside the Keplerian disk and the Compton cloud which are required for the 
Monte-Carlo simulations. In \S 4, we describe the simulation procedure and in \S 5, we present the 
results of our simulations. Finally in \S 6, we make concluding remarks.

\section{Geometry of the electron cloud and the soft photon source}

The problem at hand is very complex and thus we need to simplify the geometry of the 
inflow-outflow configuration without sacrificing the salient features. 
In Fig. 1, we present a cartoon diagram of our simulation set up. The components of the
hot electron clouds, namely, the CENBOL, the outflow and the sub-Keplerian flow, 
intercept the soft photons emerging out of the Keplerian disk 
and reprocess them via inverse Compton scattering. An injected photon may undergo a single, multiple or 
no scattering at all with the hot electrons in between its emergence from the 
Keplerian disk and its detection by the telescope at a large distance. 
The photons which enter the black holes are absorbed. The CENBOL, though toroidal in nature,
is chosen to be of spherical shape for simplicity. The sub-Keplerian inflow in the pre-shock 
region is assumed to be of wedge shape of a constant angle $\Psi$. The outflow, which 
emerges from the CENBOL in this picture is also assumed to be of constant conical angle $\Phi$.
In reality, inflow and outflow both could have somewhat different shapes, depending on the 
balance of the force components. However, the final result is not expected to be sensitive to such 
assumptions.

\begin{figure}[h]
\includegraphics[height=5.truecm,angle=0]{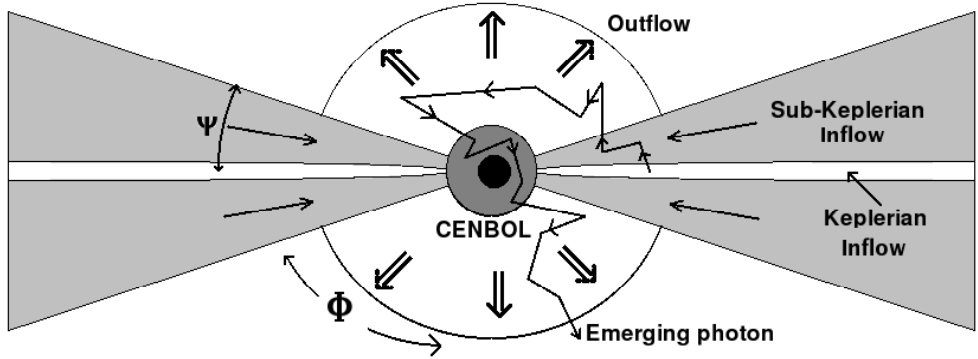}
\caption{A cartoon diagram of the geometry of our Monte-Carlo simulations
presented in this paper. The spherical inflowing post-shock region (CENBOL)
surrounds the black hole and it is surrounded by the Keplerian disk on the
equatorial plane and a sub-Keplerian halo above and below. A diverging conical outflow 
is formed from the CENBOL. Typical path of a photon is shown by zig-zag paths.
}
\end{figure}

\subsection{Distribution of temperature and density inside the Compton cloud}

We assume the black hole to be non-rotating and we use the pseudo-Newtonian
potential (Paczy\'nski \& Wiita, 1980) to describe the geometry around 
a black hole. This potential is $-\frac{1}{2(r-1)}$ (Here, $r$ is in the unit of Schwarzschild radius 
$r_g=2GM/c^2$). Velocities and angular momenta are measured are in units of 
$c$, the velocity of light and $r_g c$ respectively. 
For simplicity, we chose the Bondi accretion solution in pseudo-Newtonian 
geometry to describe both the accretion and winds. The equation of motion of the sub-Keplerian 
matter around the black hole in the steady state is assumed to be given by,
$$
u\frac{du}{dr}+\frac{1}{\rho}\frac{dP}{dr} +\frac{1}{2(r-1)^2}=0.
$$ 
Integrating this equation, we get the expression of the conserved specific energy as, 
\begin{equation}
\epsilon=\frac{u^2}{2}+na^2-\frac{1}{2(r-1)}.
\end{equation}
Here $P$ is the thermal pressure and $a$ is the adiabatic
sound speed, given by $a=\sqrt{\gamma P/\rho}$, $\gamma$ being the adiabatic
index and is equal to $\frac{4}{3}$ in our case. The conserved mass flux
equation, as obtained from the continuity equation, is given by
\begin{equation}
\dot{M}=\Omega\rho u r^2,
\end{equation}
where, $\rho$ is the density of the matter and $\Omega$ is the solid angle subtended 
by the flow. For an inflowing matter, $\Omega$ is given by, 
$$
\Omega_{in}=4\pi Sin\Psi,
$$
where, $\Psi$ is the half-angle of the conical inflow. For the outgoing matter, the 
solid angle is given by, 
$$\Omega_{out}=4\pi(1-cos\Phi),$$
where $\Phi$ is the half-angle of the conical outflow. From Eqn. 2, we get
\begin{equation}
\dot{\mu}=a^{2n}ur^2.
\end{equation}
The quantity $\dot{\mu}=\frac{\dot{m}\gamma ^n K^n}{\Omega}$ is the Chakrabarti rate
(Chakrabarti, 1989, C90, 1996) which includes the entropy, $K$ being the constant measuring the entropy of the flow,
and $n=\frac{1}{\gamma -1}$ is called the polytropic index. We take derivative
of equations (1) and (3) with respect to $r$ and eliminating $\frac{da}{dr}$ from
both the equations, we get the gradient of the velocity as, 
\begin{equation}
\frac{du}{dr}=\frac{\frac{1}{2(r-1)^2}-\frac{2a^2}{r}}{\frac{a^2}{u}-u}.
\end{equation}
From this, we obtain the Bondi accretion and wind solutions in the usual manner (C90).
Solving these equations we obtain, $u$, $a$ and finally the temperature
profile of the electron cloud ($T_e$) using $T_e=\frac{\mu a^2 m_p}{\gamma k_B}$,
where $\mu=0.5$ is the mean molecular weight, $m_p$ is the proton mass and $k_B$ is
the Boltzmann constant. Using Eq. (2), we calculate the mass density $\rho$,
and hence, the number density variation of electrons inside the Compton cloud. 
We ignore the electron-positron pair formation inside the cloud.

The flow is supersonic in the pre-shock region and sub-sonic in the post-shock (CENBOL) region. 
We chose this surface at a location where the pre-shock Mach number $M=2$.
This location depends on the specific energy $\epsilon$ (C90). 
In our simulation, we have chosen $\epsilon = 0.015$ so that we get $R_{s} = 10$.
We simulated a total of six cases. For Cases 1(a-c), we chose $\dot {m_h} = 1$, $\dot {m_d} = 0.01$ 
and for Cases 2(a-c), the values are listed in Table 2.
The velocity variation of the sub-Keplerian flow is the
inflowing Bondi solution (pre-sonic point). The density and the
temperature of this flow have been calculated according to the
above mentioned formulas. Inside the CENBOL, both the Keplerian and the
sub-Keplerian components are mixed together. The velocity variation of the
matter inside the CENBOL is assumed to be the same as the Bondi accretion flow solution
reduced by the compression ratio due to the shock. The compression ratio (i.e., 
the ratio between the post-shock and pre-shock densities) $R$ is also used 
to compute the density and the temperature profile has been calculated accordingly.  
When the outflow is adiabatic, the ratio of the outflow to the inflow rate is 
(Das et al. 2001) given by,
\begin{equation}
R_{\dot{m}} = \frac{\Omega_{out}}{\Omega_{in}} \left( \frac{f_0}{4 \gamma}\right)^3 \frac{R}{2} \left[ \frac{4}{3} \left( \frac{8(R-1)}{R^2} -1 \right) \right]^{3/2}
\end{equation}
here, we have used $n=3$ for a relativistic flow. From this, and the velocity variation obtained from 
the outflow branch of Bondi solution, we compute the density variation inside the jet. 
In our simulation, we have used $\Phi = 58^\circ$ and $\Psi = 32^\circ$. Fig. 2 shows the
variation of the percentage of matter in the outflow for these particular parameters. 

\begin{figure}[h]
\centering
\includegraphics[height=7.0truecm,angle=270]{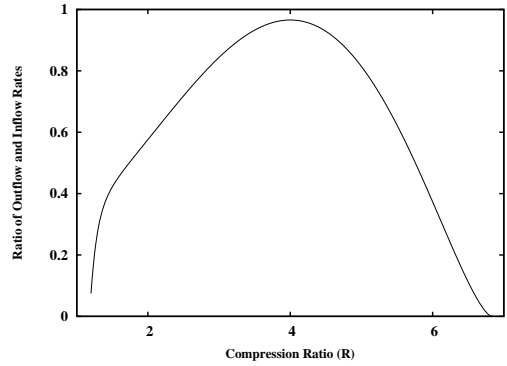}
\caption{Ratio of the outflow and the inflow rates as a function of the
compression ratio $R$ of the inflow when the outflow is adiabatic.
In our simulations, we have used the jet angle to be 58$^\circ$.}
\end{figure}

\subsection{Keplerian disk}
The soft photons are produced from a Keplerian disk whose inner edge coincides with 
CENBOL surface, while the outer edge is located at $500 r_g$.
The source of soft photons have a multi-color blackbody spectrum coming from a
standard (Shakura \& Sunyaev, 1973, hereafter SS73) disk. We assume the disk to be optically thick and the opacity
due to free-free absorption is more important than the opacity due to scattering.
The emission is black body type with the local surface temperature (SS73):
\begin{eqnarray} 
T(r) \approx 5 \times 10^7 (M_{bh})^{-1/2}(\dot{M_d}_{17})^{1/4} (2r)^{-3/4}
 \left[1- \sqrt{\frac{3}{r}}\right]^{1/4} K ,
\end{eqnarray}
The total number of photons emitted from the disk surface is obtained by integrating 
over all frequencies ($\nu$) and is given by,
\begin{eqnarray}
n_\gamma(r) = \left[16 \pi \left( \frac{k_b}{h c} \right)^3 \times 1.202057 \right]
\left(T(r)\right)^3
\end{eqnarray}
The disk between radius $r$ to $r+\delta r$ injects $dN(r)$ number of soft photons.
\begin{eqnarray}
dN(r) =  2 \pi r \delta r H(r) n_\gamma(r),
\end{eqnarray}
where, $H(r)$ is the half height of the disk given by:
\begin{eqnarray} 
H(r) = 10^5 \dot{M_d}_{17} \left[1- \sqrt{\frac{3}{r}}\right] {\rm cm}.
\end{eqnarray}
The soft photons are generated isotropically between the inner and outer edge 
of the Keplerian disk but their positions are randomized using the above distribution 
function (Eq. 8) of black body temperature $T(r)$. All the results of the simulations 
presented here have used the number of injected photons to be $6.4\times10^8$.
In the above equations, the mass of the black hole $M_{bh}$ is measured 
in units of the mass of the Sun ($M_\odot$), 
the disk accretion rate $\dot{M_d}_{17}$ is in units of $10^{17}$ gm/s. 
We chose $M_{bh} = 10$ and $\delta r = 0.5 r_g$.

\subsection{Simulation Procedure}

In a given simulation, we assume a given Keplerian rate and a given sub-Keplerian halo rate. The 
specific energy of the halo provides hydrodynamic properties (such as number density of the 
electrons and the velocity variation) and the thermal properties of matter. Since we
chose the Paczynski-Wiita (1980) potential, the radial velocity is not exactly unity at $r=1$, the horizon,
but it becomes unity just outside. In order not to over estimate the effects of bulk motion
Comptonization which is due to the momentum transfer of the moving electrons to the 
horizon, we shift the horizon just outsize $r=1$ where the velocity is unity.
The shock location of the CENBOL is chosen where the Mach number $M=2$ for simplicity and 
the compression ratio at the shock is assumed to be a free parameter.
These simplifying assumptions are not expected to affect our conclusions.
Photons are generated from the Keplerian disk according to the 
prescription in SS73 as mentioned before
and are injected into the sub-Keplerian halo, the CENBOL and the  outflowing jet.

In a simulation, we randomly generated a soft photon out of the Keplerian disk.
The energy of the soft photon at radiation temperature $T(r)$ are calculated using the 
Planck's distribution formula, where the number density of the photons 
($n_\gamma(E)$) having an energy $E$ is expressed by 
\begin{eqnarray}
n_\gamma(E) = \frac{1}{2 \zeta(3)} b^{3} E^{2}(e^{bE} -1 )^{-1},
\end{eqnarray}
where $b = 1/kT(r)$; $\zeta(3) = \sum^\infty_1{l}^{-3} = 1.202$ is the Riemann zeta function.

Using another set of random numbers we obtained the direction of the injected photons and with yet
another random number we obtained a target optical depth $\tau_c$ at which the scattering takes place.
The photon was followed within the CENBOL till the optical
depth ($\tau$) reached $\tau_c$. The increase in optical depth ($d\tau$) during its traveling  of 
a path of length $dl$ inside the electron cloud is given by: $d\tau = \rho_n \sigma dl$, where 
$\rho_n$ is the electron number density.

The total scattering cross section $\sigma$ is given by Klein-Nishina formula:
\begin{equation}
\sigma = \frac{2\pi r_{e}^{2}}{x} \\
\left[ \left( 1 - \frac{4}{x} - \frac{8}{x^2} \right) ln\left( 1 + x \right) + \frac{1}{2} + \frac{8}{x} - \frac{1}{2\left( 1 + x \right)^2} \right],
\end{equation}
where, $x$ is given by,
\begin{equation}
x = \frac{2E}{m c^2} \gamma \left(1 - \mu \frac{v}{c} \right),
\end{equation}
$r_{e} = e^2/mc^2$ is the classical electron radius and $m$ is the mass of the electron.

We have assumed here that a photon of energy $E$ and momentum $\frac{E}{c}\bf{\widehat{\Omega}}$
is scattered by an electron of energy $\gamma mc^{2}$ and momentum $\overrightarrow{\bf{p}} = \gamma m \overrightarrow{\bf{v}}$, with $\gamma = \left( 1 - \frac{v^2}{c^2}\right)^{-1/2}$ and $\mu = \bf{\widehat{\Omega}}. \widehat{\bf{v}}$.
At this point a scattering is allowed to take place. The photon selects an electron and the energy
exchange is computed through Compton or inverse Compton scattering formula. The electrons
are assumed to obey relativistic Maxwell distribution inside the CENBOL.
The number $dN(p)$ of Maxwellian electrons having momentum between 
$\vec{p}$ to $\vec{p} + d\vec{p}$ is expressed by, 
\begin{eqnarray}
dN(\vec{p}) = exp[-(p^2c^2 + m^2c^4)^{1/2}/kT_e]d\vec{p}.
\end{eqnarray}
Generally, the same procedure as in Paper I was used, except that we are now focusing on those photons also
photons which were scattered at least once by the outflow. We are especially choosing the
cases when the jet could play a major role in shaping the spectrum.

\section{Results and Discussions}

In Fig. 3(a-c) we present the velocity, electron number density and temperature variations as a function of the 
radial distance from the black hole for specific energy $\epsilon=0.015$. $\dot{m_d} = 0.01$ and 
$\dot{m_h} = 1$ were chosen. Three cases were run by varying the compression ratio $R$. These 
are given in Col. 2 of Table 1. The corresponding percentage of matter going in the 
outflow is also given in Col. 2. In the left panel, the bulk velocity variation is shown.
The solid, dotted and dashed curves are the same for $R = 2$ (Case 1a), $4$ 
(Case 1b) and $6$ (Case 1c) respectively. The same line style is used in other panels. The velocity 
variation within the jet does not change with $R$, but the density (in the 
unit of $cm^{-3}$) does (middle panel). 
The doubledot-dashed line gives the velocity variation of the matter within the 
jet for all the above cases. The arrows show the direction of the bulk velocity 
(radial direction in accretion, vertical direction in jets). The last panel gives the temperature (in keV) 
of the electron cloud in the CENBOL, jet, sub-Keplerian and Keplerian 
disk. Big dash-dotted line gives the temperature profile inside the Keplerian disk.

\begin{figure}[h]
\centering
\includegraphics[height=14truecm,angle=270]{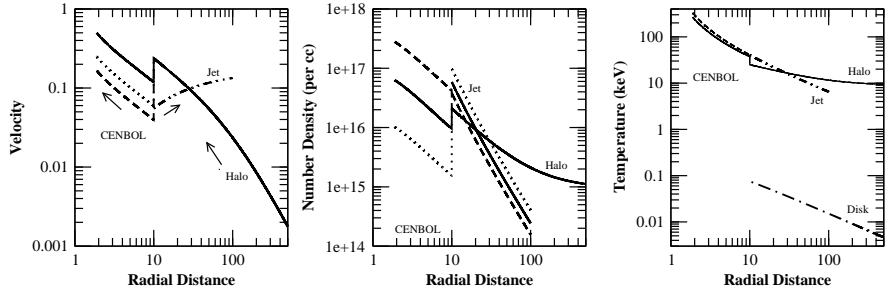}
\vskip -4.0cm
\caption{(a-c): Velocity (left), density (middle) and the temperature (right) profiles
of Cases 1(a-c) as described in Table 1 
are shown with solid ($R=2$), dotted ($R=4$) and dashed ($R=6$) curves. 
$\dot{m_d} = 0.01$ and $\dot{m_h} = 1$ were used.}
\end{figure}

\begin{figure}[h]
\centering
\includegraphics[height=15truecm,angle=270]{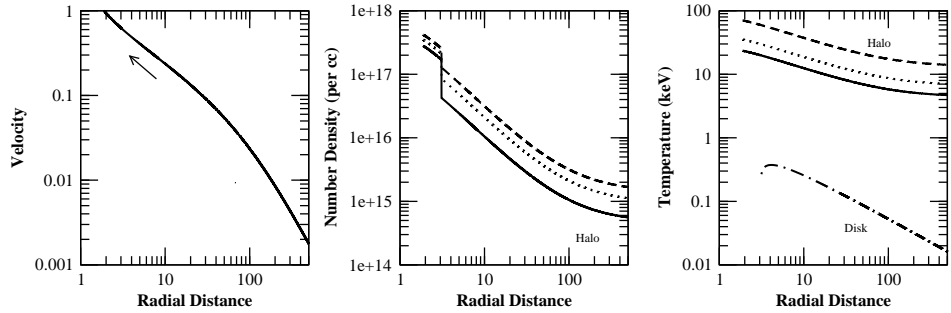}
\vskip -4.0cm
\caption{{\bf (a-c)}: 
Velocity (left), density (middle) and the temperature (right) profiles of Cases 2(a-c) as described in Table 2  
are shown with solid ($\dot{m_h} = 0.5$), dotted ($1$) and dashed ($1.5$)
curves. $\dot{m_d} = 1.5$ was used throughout. Velocities are the 
same for all the disk accretion rates.}
\end{figure}

In Figs. 4(a-c), we show the velocity (left), number density of electrons 
(middle) and temperature (right) profiles
of Cases 2(a-c) as described in Table 2. Here we have fixed $\dot{m_d}=1.5$ and
$\dot{m_h}$ is varied: ${\dot {m_h}}=\ 0.5$ (solid), $1$ (dotted) and $1.5$ (dashed).
No jet is present in this case ($R=1$). To study the effects of bulk
motion Comptonization, the temperature of the electron cloud has been kept low for these cases. The
temperature profile in the different cases has been chosen according to the Fig. 3b of CT95.
The temperature profile of the Keplerian disk for the above cases has been marked as `Disk' .

{\tiny
\begin {tabular}[h]{|c|c|c|c|c|c|c|c|c|c|}
\hline
\multicolumn{10}{|c|}{Table 1}\\
\hline Case & R, $P_m$ & $N_{int}$ & $N_{cs}$ & $N_{cenbol}$ & $N_{jet}$ & $N_{subkep}$ & $N_{cap}$ & $p$ & $\alpha$\\
\hline 1a & 2, 58  & 2.7E+08 & 4.03E+08 & 1.35E+07 & 7.48E+07  & 8.39E+08 & 3.35E+05  & 63 & 0.43 \\
\hline 1b & 4, 97 & 2.7E+08 & 4.14E+08  & 2.39E+06 & 1.28E+08 & 8.58E+08 & 3.27E+05 & 65 & 1.05  \\
\hline 1c & 6, 37 & 2.7E+08 & 3.98E+08 & 5.35E+07 & 4.75E+07 & 8.26E+08 & 3.07E+05 &  62 & -0.4 \\
\hline
\end{tabular}
}

In Table 1, we summarize the details of all the Cases results of which were depicted in Fig. 3(a-c).
In Col. 1, various Cases are marked. In Col. 2, the compression ratio ($R$) and 
percentage $P_m$ of the total matter that is going out as outflow (see, 
Fig. 2) are listed. In Col. 3, we show the total number of photons
(out of the total injection of $6.4 \times 10^8$)
intercepted by the CENBOL and jet ($N_{int}$) combined. Column 4 gives
the number of photons ($N_{cs}$) that have suffered Compton scattering 
inside the flow. Columns 5, 6 and 7 show the number of
scatterings which took place in the CENBOL ($N_{cenbol}$), in the jet ($N_{jet}$) and in
the pre-shock sub-Keplerian halo ($N_{subkep}$) respectively. A comparison of them will give
the relative importance of these three sub-components of the sub-Keplerian disk. The number
of photons captured ($N_{cap}$) by the black hole is given
in Col. 8. In Col. 9, we give the percentage $p$ of the total 
injected photons that have suffered scattering through CENBOL and the jet. 
In Col. 10, we present the energy spectral index
$\alpha$ ($I(E) \sim E^{-\alpha}$) obtained from our simulations.

{\tiny
\begin {tabular}[h]{|c|c|c|c|c|c|c|c|c|}
\hline
\multicolumn{9}{|c|}{Table 2}\\
\hline Case & $\dot{m_h}$, $\dot{m_d}$ & $N_{int}$ & $N_{cs}$ & $N_{ms}$ &  
$N_{subkep}$ & $N_{cap}$ & $p$ & $\alpha_1, \alpha_2$\\
\hline 2a & 0.5, 1.5 & 1.08E+06  & 2.13E+08 & 7.41E+05 & 3.13E+08 & 1.66E+05 & 33.34 & -0.09, 0.4\\
\hline 2b & 1.0, 1.5 & 1.22E+06  & 3.37E+08 & 1.01E+06 & 6.82E+08 & 2.03E+05 & 52.72  & -0.13, 0.75\\
\hline 2c & 1.5, 1.5 & 1.34E+06  & 4.15E+08 & 1.26E+06 & 1.11E+09 & 2.29E+05 & 64.87  & -0.13, 1.3\\
\hline
\end{tabular}
}

In Table 2, we summarize the results of simulations where we have varied $\dot{m_d}$, 
for a fixed value of $\dot{m_h}$. In all of these cases no jet comes out of the CENBOL 
(i.e., $R=1$). In the last column, we listed two spectral slopes $\alpha_1$ (from $10$ to 
$100$keV) and $\alpha_2$ (due to the bulk motion Comptonization). 
Here, $N_{ms}$ represents the photons that have suffered scattering between $r_g=3$ and the
horizon of the black hole.

In Fig. 5, we show the variation of the spectrum in the three simulations presented in Fig. 3(a-c).
The dashed, dash-dotted and doubledot-dashed lines are for $R=2$ (Case 1a), $R=4$ (Case 1b) 
and $R=6$ (Case 1c) respectively. The solid curve gives the spectrum of
the injected photons. Since the density, velocity and
temperature profiles of the pre-shock, sub-Keplerian region and the
Keplerian flow are the same in all these cases, we find that
the difference in the spectrum is mainly due to the
CENBOL and the jet. In the case of the strongest shock
(compression ratio $R=6$), only $37\%$ of the total injected
matter goes out as the jet. At the same time, due
to the shock, the density of the post-shock region increases
by a factor of $6$. Out of the three cases, the effective density of the
matter inside CENBOL is the highest and that inside the jet is
the lowest in this case. Again, due to the shock, the temperature
increases inside the CENBOL and hence the spectrum is the hardest.
Similar effects are seen for moderate shock ($R=4$) and 
to a lesser extent, the low strength shock ($R=2$) also. 
When $R=4$, the density of the post-shock region
increases by the factor of $4$ while almost $97\%$ of total injected
matter (Fig. 2) goes out as the jet reducing the matter density
of the CENBOL significantly. From
Table 1 we find that the $N_{cenbol}$ is the lowest and
$N_{jet}$ is the highest in this case (Case 1b). This
decreases the up-scattering and increases the down-scattering
of the photons. This explains why the spectrum is
the softest in this case. In the case of low strength shock
($R=2$), $57\%$ of the inflowing matter goes out as jet, but due
to the shock the density increases by factor of $2$ in the
post-shock region. This makes the density similar to a non-shock
case as far as the density is concerned, but with a little higher
temperature of the CENBOL due to the shock. So the spectrum with
the shock would be harder than when the shock is not present.
The disk and the halo accretion rates used for these cases
are $\dot{m_d} = 0.01$ and $\dot{m_h} = 1$.
 
\begin{figure}[h]
\centering
\includegraphics[height=10truecm,angle=270]{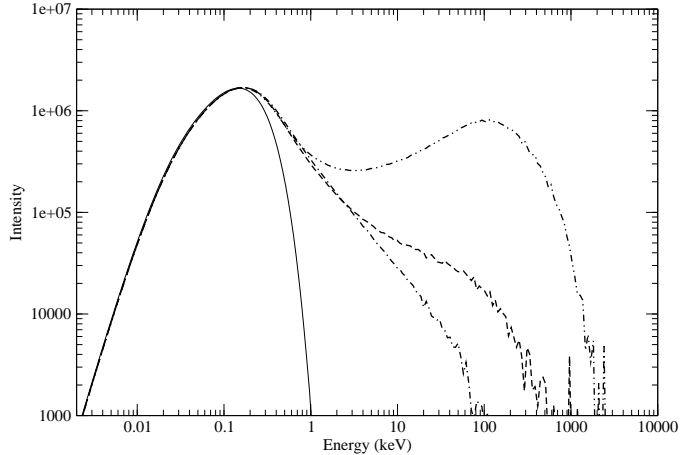}
\caption{Variation of the emerging spectrum for different compression ratios.
The solid curve is the injected spectrum from the Keplerian disk. The dashed, dash-dotted 
and doubledot-dashed lines are for $R=2$ (Case 1a), $R=4$ (Case 1b) and $R=6$ (Case 1c) 
respectively. The disk and halo accretion rates used for these cases are $\dot{m_d} 
= 0.01$ and $\dot{m_h} = 1$. See, text for details.}
\end{figure}

In Fig. 6, we show the components of the emerging spectrum for all the three cases 
presented in Fig. 5. The solid curve is the intensity of all the photons
which suffered at least one scattering. The dashed curve corresponds to the photons emerging
from the CENBOL region and the dash-dotted curve is for the photons
coming out of the jet region. We find that the spectrum from
the jet region is softer than the spectrum from the CENBOL. As $N_{jet}$
increases and  $N_{cenbol}$ decreases, the spectrum from the jet becomes
softer because of two reasons. First, the temperature of the jet is lesser
than that of the CENBOL, so the photons get lesser amount of energy from thermal
Comptonization making the spectrum softer. Second, the photons are
down-scattered by the outflowing jet which eventually make the spectrum
softer. We note that a larger number of photons are present in the spectrum 
from the jet than the spectrum from the CENBOL, which shows the photons
have actually been down-scattered. The effect of down-scattering is larger when $R=4$. 
For $R=2$ also there is significant amount of 
down scattered photons. But this number is very small for the case $R=6$ 
as $N_{cenbol}$ is much larger than $N_{jet}$ so most of the photons get 
up-scattered. The difference between total (solid) and the sum of the 
other two regions gives an idea of the contribution from the sub-Keplerian
halo located in the pre-shock region. In our choice of geometry (half angles of the
disk and the jet), the contribution of the pre-shock flow is significant. In general
it could be much less. This is especially true when the CENBOL is farther out.

\begin{figure}[h]
\centering
\includegraphics[height=15truecm,angle=270]{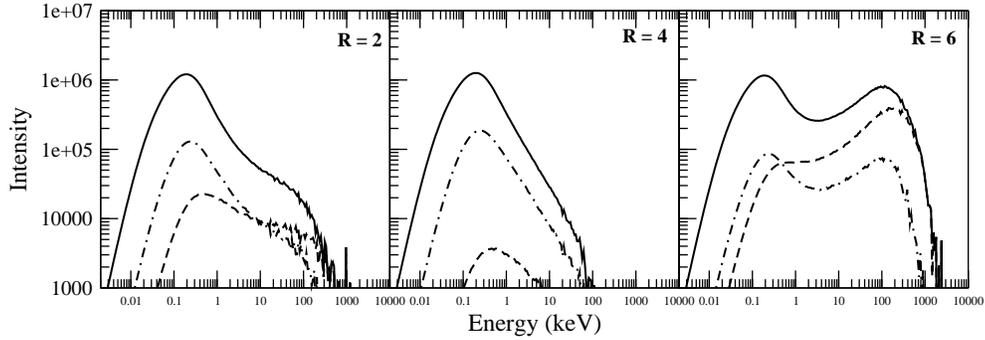}
\vskip -4.0cm
\caption{{\bf (a-c)}: Variation of the components of the emerging spectrum with 
the shock strength (R). The dashed curves correspond to the photons emerging
from the CENBOL region and the dash-dotted curves are for the photons 
coming out of the jet region. The solid curve is the spectrum for all 
the photons that have suffered scatterings. See, the text for details. }
\end{figure}

In Fig. 7, the emerging spectra due to the bulk motion 
Comptonization when the halo rate is varied. The solid curve is the injected
spectrum (modified black body). The dotted, dashed, and dash-dotted curves 
are for $\dot{m_h} = 0.5, \ 1$ and $1.5$ respectively. $\dot{m_d} = 1.5$ 
for all the cases. The Keplerian disk extends up to $3 r_g$. Table 2 
summarizes the parameter used and the results of the simulation.
As the halo rate increases, the density of the CENBOL also
increases causing a larger number of scattering. 
From Fig. 4a, we noticed that the bulk velocity variation of the electron cloud is the same for
all the four cases. Hence, the case where the density is
maximum, the photons got energized to a very high value due
to repeated scatterings with that high velocity cold matter.
As a result, there is a hump in the spectrum around 100 keV energy
for all the cases. We find the signature of two power-law regions in the higher energy
part of the spectrum. The spectral indices are  given in Table 2. It
is to be  noted that $\alpha_2$ increases with $\dot{m_h}$ and becomes softer for high  $\dot{m_h}$.
Our geometry here at the inner edge is conical which is more realistic, 
unlike a sphere (perhaps nonphysically so)
in Laurent \& Titarchuk (2001). This may be the reason why our slope is not the 
same as in Laurent \& Titarchuk (2001) where $\alpha_2=1.9$. In Fig. 8,  
we present the components of the emerging spectra.
As in Fig. 6, solid curves are the spectra of all the photons that have suffered
scattering. The dashed and dash-dotted curves are the spectra
of photons emitted from inside and outside of the 
marginally stable orbit ($3 r_g$) respectively. The photons from 
inside the marginally stable radius are Comptonized by the 
bulk motion of the converging infalling matter and produces the power-law tail 
whose spectral index is given by $\alpha_2$ (Table 2).

\begin{figure}[h]
\centering
\includegraphics[height=12truecm,angle=270]{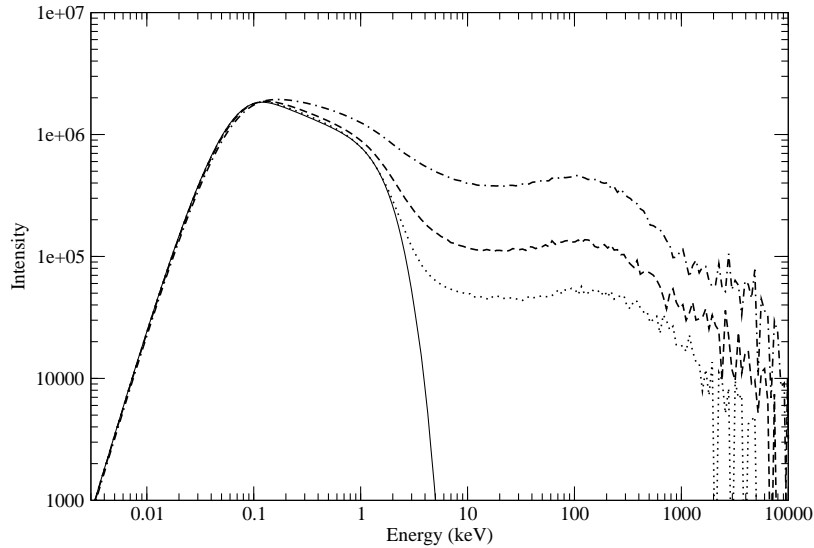}
\caption{Bulk motion Comptonization spectrum.
Solid (Injected), dotted ($\dot{M_h} = 0.5$), dashed ($\dot{M_h} = 1$), 
dash-dotted ($\dot{M_h} = 1.5$). $\dot{M_d} = 1.5$ for all the cases.
Keplerian disk extends up to $3.1 r_g$. Table 2 summarizes the 
parameters used and the simulation results for these cases.}
\end{figure}

\begin{figure}[h]
\centering
\includegraphics[height=14truecm,angle=270]{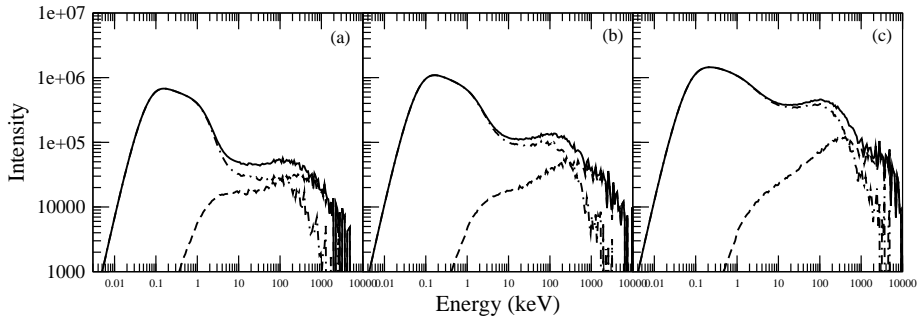}
\vskip -4.0cm
\caption{Components of the emerging spectrum for the Cases 2(a-c).
Solid curves are the spectra of all the photons that have suffered
scattering. The dashed and dash-dotted curves are the spectra
of photons which are emitted from inside and outside of the 
marginally stable orbit ($3 r_g$) respectively. The photons from 
inside the marginally stable radius are Comptonized by the 
bulk motion of the infalling matter. Here the jet is absent.}
\end{figure}

\section{Concluding remarks}

In this paper, we extended the results of our previous work on Monte-Carlo simulations
(Paper I). We included the outflow in conjunction with the inflow. The outflow rate
was self-consistently computed from the inflow rate using well-known considerations
present in the literature (Das et al. 2001 and references therein). 
We compute the effects of the thermal and the bulk motion Comptonization 
on the soft photons emitted from a Keplerian disk around a black hole by the post-shock region
of a sub-Keplerian flow which surrounds the Keplerian disk. A shock in the inflow increases the 
CENBOL temperature, increases the electron number density and reduces the bulk velocity. 
Thermal Comptonization and bulk motion Comptonization inside the CENBOL increases photon energy. 
However, the CENBOL also generates the outflow of matter which 
down-scatters the photons to lower energy. We show that the thermal Comptonization and the bulk motion 
Comptonization were possible by both the accretion and the outflows. While the converging flow
up-scatters the radiation, the outflow down-scatters. However, the net effect is not simple.
The outflow parameters are strongly coupled to the inflow parameters and thus for a given
inflow and outflow geometry,  the strength of 
the shock can also determine whether the net scattering by the jets would be significant or not.
Sometimes the spectrum may become very complex with two power-law indices, one from 
thermal and the other from the bulk motion Comptonization. Since the volume of the 
jet may be larger than that of the CENBOL, sometimes the number of scatterings suffered by 
softer photons from the electrons in the jet may be high. 
However, whether the CENBOL or the jet emerging from it will dominate in shaping the 
spectrum strongly depends on the geometry of the flow and the strength of the shock.
We also found that the halo can Comptonize and harden the spectrum even without the CENBOL.

\section*{Acknowledgments}

The work of HG is supported by a RESPOND project.

\end{document}